 \documentclass[final,5p,times,twocolumn]{elsarticle}

\usepackage{hyperref}
\usepackage{amssymb}
\usepackage{setspace}
\usepackage[version=4]{mhchem}

\usepackage{lineno}

\journal{Nucl. Inst. and Methods in Physics Research B}

\begin{document}

\begin{frontmatter}



\title{In-source and in-trap formation of molecular ions in the actinide mass range at CERN-ISOLDE}


\author[cern,jgu]{M. Au\corref{cor1}}
\ead{mia.au@cern.ch}

\author[cern,kul]{M. Athanasakis-Kaklamanakis}

\author[cern,greifswald]{L. Nies}

\author[cern,frib]{J. Ballof}

\author[cern,marburg]{R. Berger}

\author[cern]{K. Chrysalidis}

\author[greifswald]{P. Fischer}

\author[cern]{R. Heinke}

\author[kul]{J. Johnson}

\author[cern,ill]{U. K{\"o}ster}

\author[cern]{D. Leimbach\fnref{ugot}}

\author[cern]{B. Marsh}

\author[cern,mpik]{M. Mougeot\fnref{jyfl}}

\author[manchester]{J. Reilly}

\author[cern]{E. Reis}

\author[darmstadt]{M. Schlaich}

\author[cern,mpik]{Ch. Schweiger}

\author[greifswald]{L. Schweikhard}

\author[cern]{S. Stegemann}

\author[cern,manchester]{J. Wessolek}

\author[darmstadt]{F. Wienholtz}

\author[cern,mit]{S. G. Wilkins}

\author[kul]{W. Wojtaczka}

\author[jgu,gsi,him]{Ch. E. D{\"u}llmann}

\author[cern]{S. Rothe}

\cortext[cor1]{Corresponding author}
\fntext[ugot]{Present address: University of Gothenburg, Sweden}
\fntext[jyfl]{Present address: University of Jyv{\"a}skyl{\"a}, Finland}

\address[cern]{CERN, Geneva, Switzerland}
\address[jgu]{Johannes Gutenberg-Universit{\"a}t Mainz, Mainz, Germany}
\address[kul]{KU Leuven, Leuven, Belgium}
\address[greifswald]{Universit{\"a}t Greifswald, Greifswald, Germany}
\address[frib]{FRIB, Michigan State University, East Lansing, Michigan}
\address[marburg]{Philipps-Universit{\"a}t Marburg, Marburg, Germany}
\address[ill]{Institut Laue-Langevin, Grenoble, France}
\address[mpik]{Max Planck Institut f{\"u}r Kernphysik, Heidelberg, Germany}
\address[manchester]{The University of Manchester, Manchester, United Kingdom}
\address[darmstadt]{Technische Universit{\"a}t Darmstadt, Darmstadt, Germany}
\address[mit]{Massachusetts Institute of Technology, Cambridge, United States of America}
\address[gsi]{GSI Helmholtzzentrum f{\"u}r Schwerionenforschung, Darmstadt, Germany}
\address[him]{Helmholtz Institute Mainz, Mainz, Germany}

\begin{abstract}
The use of radioactive molecules for fundamental physics research is a developing interdisciplinary field limited dominantly by their scarce availability. In this work, radioactive molecular ion beams containing actinide nuclei extracted from uranium carbide targets are produced via the Isotope Separation On-Line technique at the CERN-ISOLDE facility. Two methods of molecular beam production are studied: extraction of molecular ion beams from the ion source, and formation of molecular ions from the mass-separated ion beam in a gas-filled radio-frequency quadrupole ion trap. Ion currents of U$^+$, \ce{UO_{1-3}^+}, \ce{UC_{1-3}}, \ce{UF_{1-4}^+}, \ce{UF_{1,2}O_{1,2}^+} are reported. Metastable tantalum and uranium fluoride molecular ions are identified. Formation of \ce{UO_{1-3}^+}, \ce{U(OH)_{1-3}^+}, \ce{UC_{1-3}}, \ce{UF_{1,2}O_{1,2}^+} from mass-separated beams of U$^+$, \ce{UF_{1,2}^+} with residual gas is observed in the ion trap. The effect of trapping time on molecular formation is presented. 
\end{abstract}
\end{frontmatter}
\section{Introduction}
\label{sec:intro}
There is interdisciplinary interest in radioactive molecules bridging fields of molecular physics, atomic physics and nuclear physics, as well as physics beyond the standard model \cite{radmols2023}. Experimental research possibilities with many radioactive molecules are currently constrained by their limited production. This is particularly the case for radioactive molecules containing an actinide element. Only actinides in the decay chains of primordial $^{232}$Th and $^{235,238}$U are available in macroscopic quantities in nature. All others must be produced artificially. 

The Isotope Separation On-Line (ISOL) method allows production of a wide range of radioactive nuclides across the nuclear chart through reactions induced by the impact of an accelerated particle beam hitting a thick target. The ISOLDE facility at CERN \cite{Catherall2017} uses 1.4-GeV protons accelerated by CERN’s Proton Synchrotron Booster (PSB) and can employ a variety of target and ion source systems. Once created, the reaction products must diffuse out of the target material and effuse to the ion source, where they are ionized and extracted as a beam of charged particles. For refractory species, forming volatile compounds has been employed as a technique to improve extraction from the target by delivering the isotopes of interest as molecular ion beams \cite{Ballof2019, Koster2008, Koster2007, Eder1992}. In specific cases, the formation of molecules can reduce the isobaric contamination remaining after mass separation. The production of actinide molecules could address the scarcity and purity problems limiting many experiments on actinide isotopes. In addition, they present promising cases themselves \cite{radmols2023,Skripnikov2020,Safronova2018,Isaev2010}. 

\section{Method}
\label{sec:method}

\begin{figure}
	\includegraphics[width=0.5\textwidth]{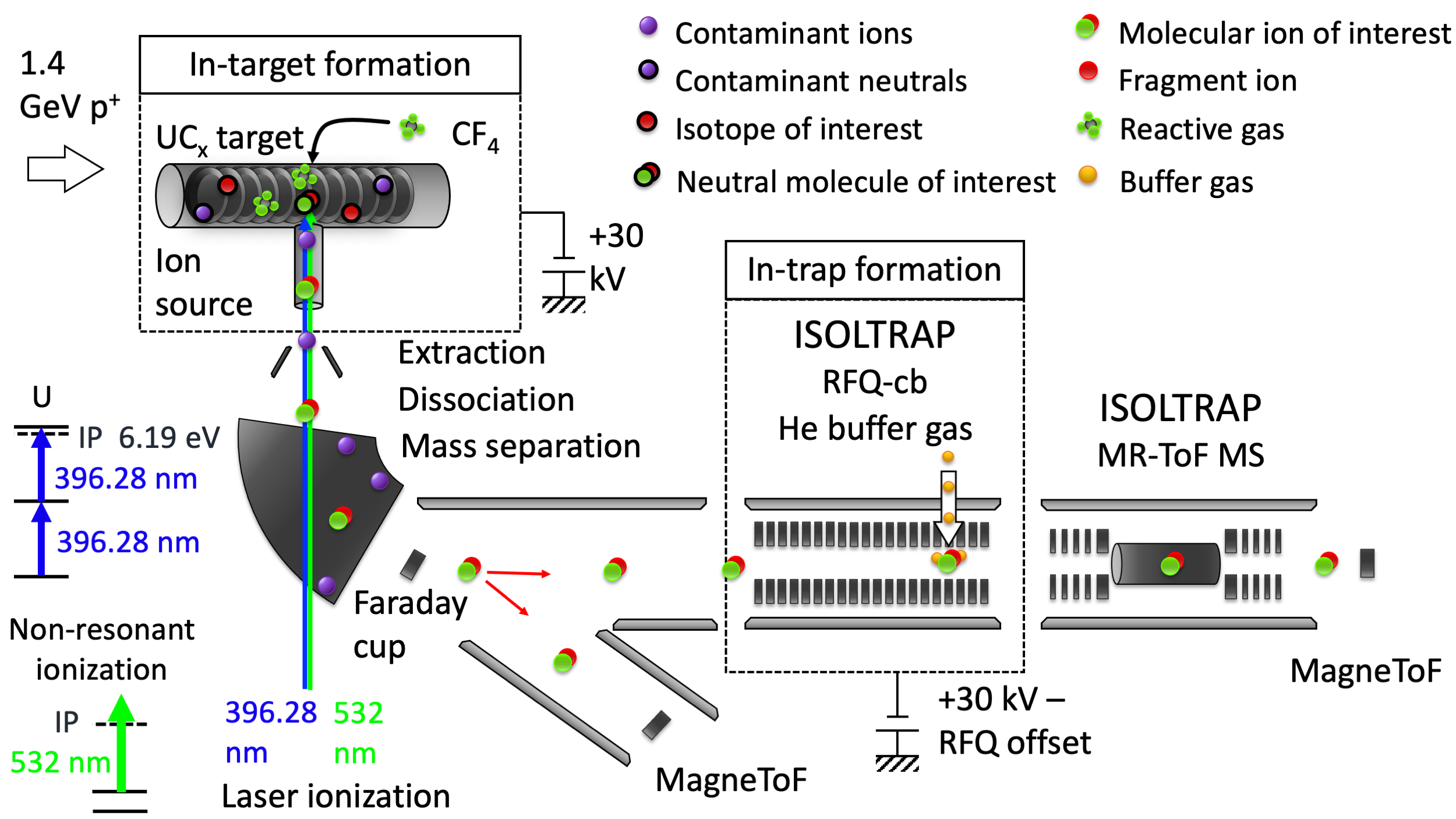}%
	\caption{Schematic of the experimental setup. Molecules are generated in the UC$_\textrm{x}$ target or from the mass-separated ion beam in the ISOLTRAP RFQ-cb. The mass-separated beam is sent to a MagneToF detector or bunched and sent to the MR-ToF MS for identification. The uranium resonant ionization scheme shown uses a single titanium:sapphire laser \cite{Savina2021}.\label{fig:schematic}}
\end{figure}

The ISOLDE facility was used to study actinide species produced from four porous micro-structured uranium carbide (UC$_\textrm{x}$) target units: a previously-irradiated target coupled to a rhenium surface ion source; a previously-irradiated target coupled to a tungsten surface ion source; and two new targets coupled to Forced Electron Beam Induced Arc Discharge (FEBIAD) ion sources \cite{Penescu2010a}. The ISOLDE Resonance Ionization Laser Ion Source (RILIS \cite{Fedosseev2017}) was used to resonantly ionize atomic U with the ionization scheme shown in Fig. \ref{fig:schematic}. Ion beams were extracted from the ion source using a 30-kV potential difference and separated by their mass-to-charge ratio in the separator magnet. Mass-separated ion beams were either sent to a MagneToF detector or cooled and bunched in the ISOLTRAP Radio-Frequency Quadrupole cooler-buncher (RFQ-cb) \cite{HERFURTH2001}. The bunched beam was sent to the Multi-Reflection Time-of-Flight Mass Spectrometer (MR-ToF MS) \cite{Wolf2013}, where ions were separated based on their mass-to-charge ratios, including isobars, which were identified through ToF mass measurements. The experimental setup is shown schematically in Figure \ref{fig:schematic}. 

\section{In-source molecular formation}
\label{sec:in-source}
\begin{figure}
    \includegraphics[width=0.5\textwidth]{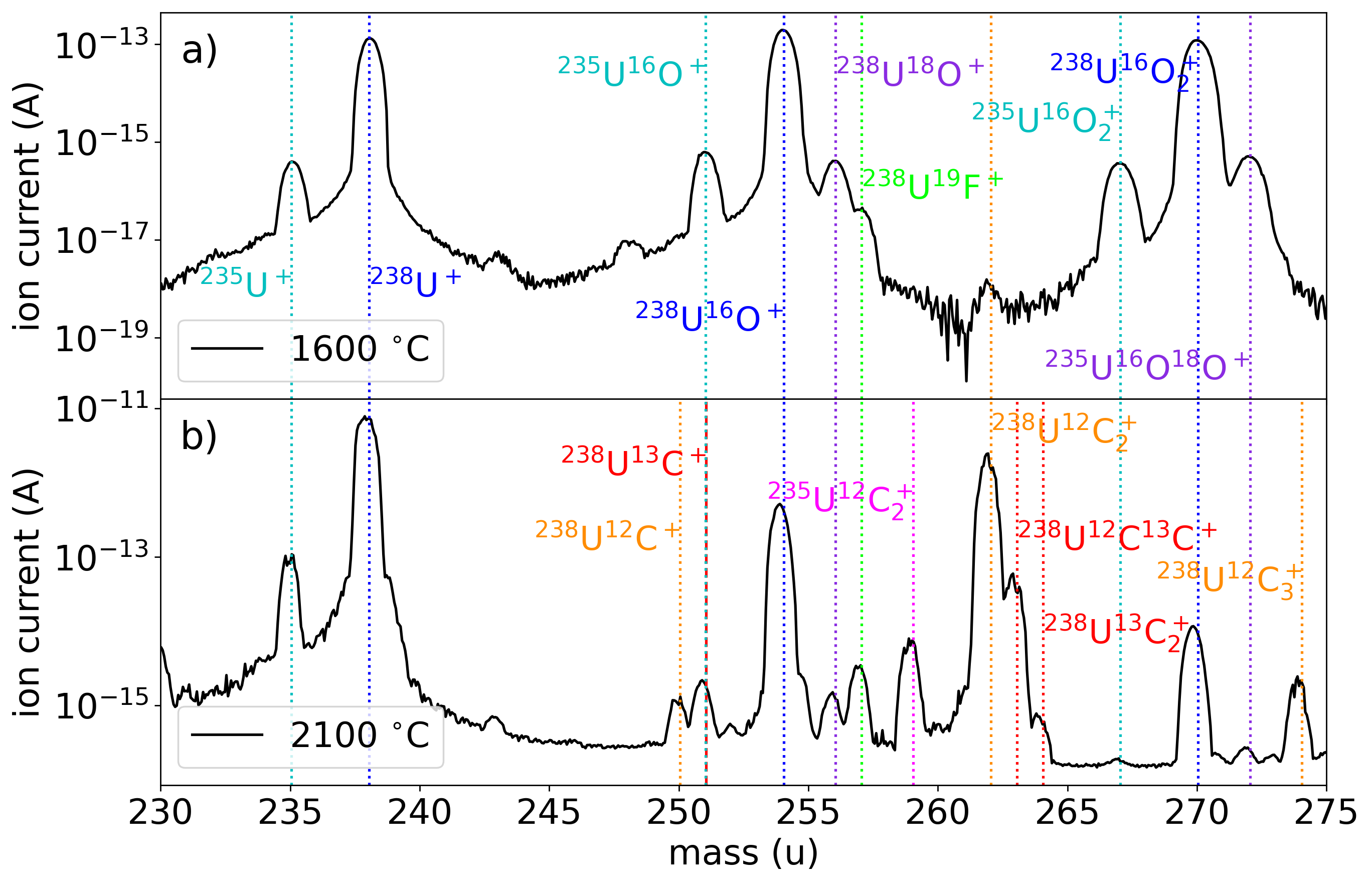}
	\caption{Ion beam current recorded on the MagneToF detector during a scan of the GPS mass separator magnet showing the surface-ionized molecules from a rhenium ion source coupled to a previously irradiated UC$_x$ target at a target temperature of a) 1600\,$^{\circ}$C and b) 2100\,$^{\circ}$C. Note the logarithmic scale. \label{fig:mass_scan_pre-rad}}
\end{figure}

The target units with the tungsten surface ion source and the two FEBIAD type ion sources were equipped with calibrated leaks (1.3E-4, 3E-4 and 5.7E-5\,mbar\,L\,s$^{-1}$) through which carbon tetrafluoride (carbon tetrafluoromethane, CF$_4$) gas was injected as a reagent for fluoride molecule formation. Using the two different types of ion sources, surface-, electron-impact-, and non-resonantly laser-ionized-molecules were observed. Experimental parameters are indicated in the captions of Figures ~\ref{fig:mass_scan_pre-rad} and \ref{fig:surface_vs_plasma}. 

Uranium molecules from the target material were identified using mass scans performed with the ISOLDE mass separator magnets (Fig.~\ref{fig:mass_scan_pre-rad}) and verified using the ISOLTRAP MR-ToF MS. $^{235,238}$U are present in the target as well as trace amounts of $^{234}$U as a product of $^{238}$U decay. These formed UO$^+$ and UO$_2^+$ with $^{16,18}$O present from residual gas or oxide residues in the target. Depending on the degree of oxidation, the molecular oxide ion intensity decreased with time and target heating in the presence of excess carbon from the UC$_\textrm{x}$ target. At higher temperatures, uranium carbides (UC$^+$ and UC$_2^+$) were observed, formed from both $^{12,13}$C (Fig. \ref{fig:mass_scan_pre-rad}). With the addition of CF$_4$, UF$_{1,2,3}^+$ and UFO$^+$ dominated the total surface- or plasma-ionized beam. Details of experimental parameters are given in Fig.~\ref{fig:surface_vs_plasma}. 

\subsection{Non-resonant laser and plasma ionization}

\begin{figure}
	\includegraphics[width=0.5\textwidth]{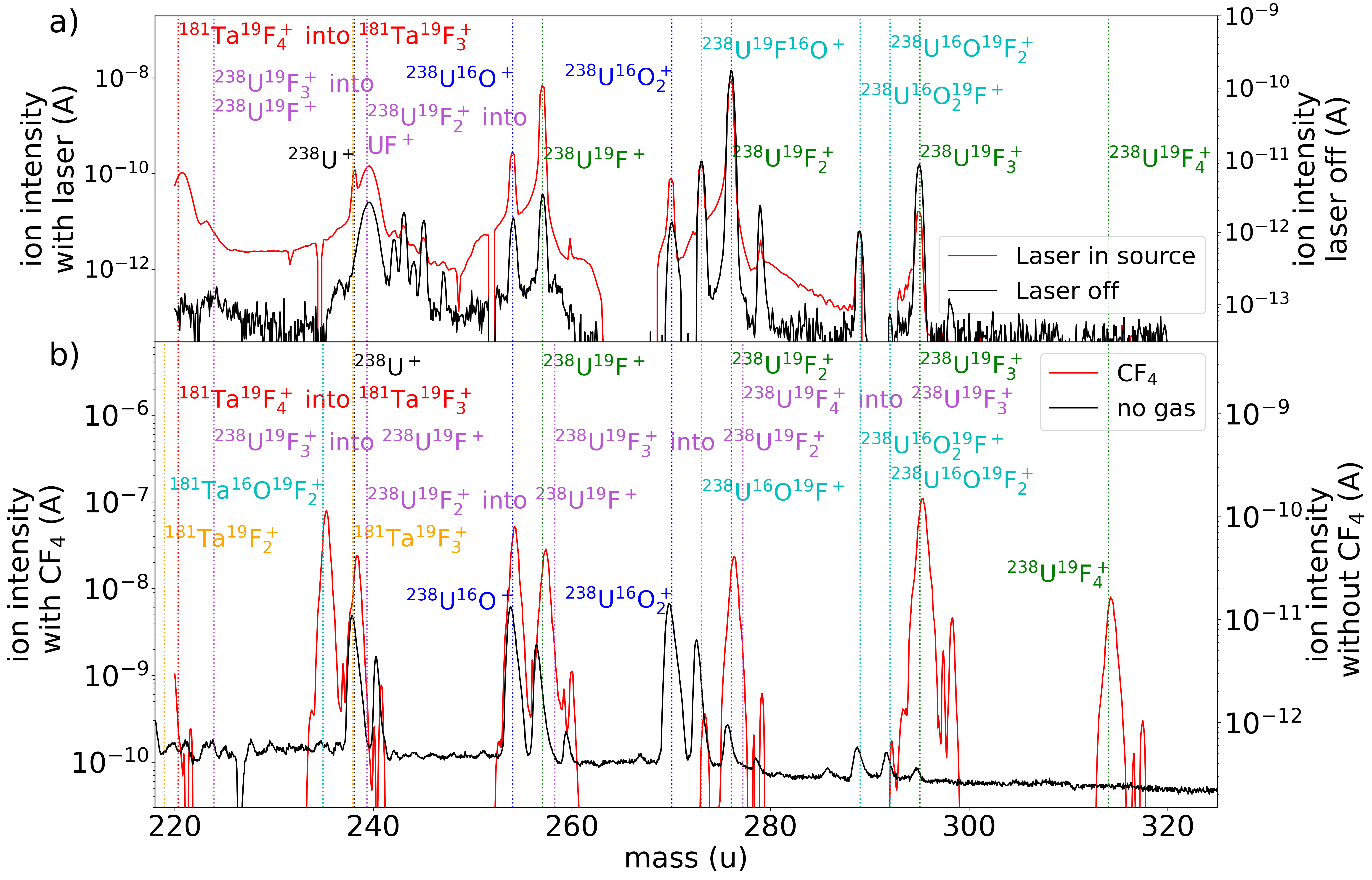}%
	\caption{Mass spectra of ion beams from UC$_{\textrm{x}}$ targets using the GPS mass separator magnet. a) a tungsten surface ion source without (black) and with 38\,W of 532-nm laser light at a repetition rate of 10\,kHz (red). Target at 1650\,$^{\circ}$C and ion source at 1370\,$^{\circ}$C. b) a FEBIAD ion source before (black) and after injection of CF$_4$ through a calibrated leak for an ion source internal pressure of approximately 7.5E-5 mbar (red). Target at 1200\,$^{\circ}$C and ion source at 1830\,$^{\circ}$C. \label{fig:surface_vs_plasma}}
\end{figure}

UO$^+$ and UO$_2^+$ dominate the ion beam for oxidized targets. With first ionization potentials of 6.0313(6)\,eV and 6.128(3)\,eV for UO and UO$_2$, respectively  \cite{Morss2010}, these species are observed with both surface and FEBIAD ion sources. With CF$_4$ injection, tungsten surface ionization, and 30\,W of 532-nm laser light, UF$^+$ and UF$_2^+$ are the most intense uranium molecular ion beams. Surface-ionized UF$_3^+$ is detectable with a Faraday Cup; UF$_4^+$ is not observed (Fig.~\ref{fig:surface_vs_plasma} a). Using a FEBIAD ion source, the UF$_3^+$ sideband is dominant and UF$_4^+$ is observed. Higher rates of U$^+$ likely result from the breakup of uranium molecules in the FEBIAD ion source before extraction as an ion beam. Sideband ratios depend strongly on the concentration of CF$_4$, favouring UF$_{2,3}^+$ with higher leak rates of CF$_4$. Bond dissociation energies of UO (7.856(135)\,eV), UO$_2$ (7.773(145)\,eV) \cite{Morss2010} suggest that some dissociation of neutral and singly-charged oxides should occur within the FEBIAD ion source. 

\subsection{Metastable molecular ions}
\label{sec:metastable}

\begin{figure}
	\includegraphics[width=0.4\textwidth]{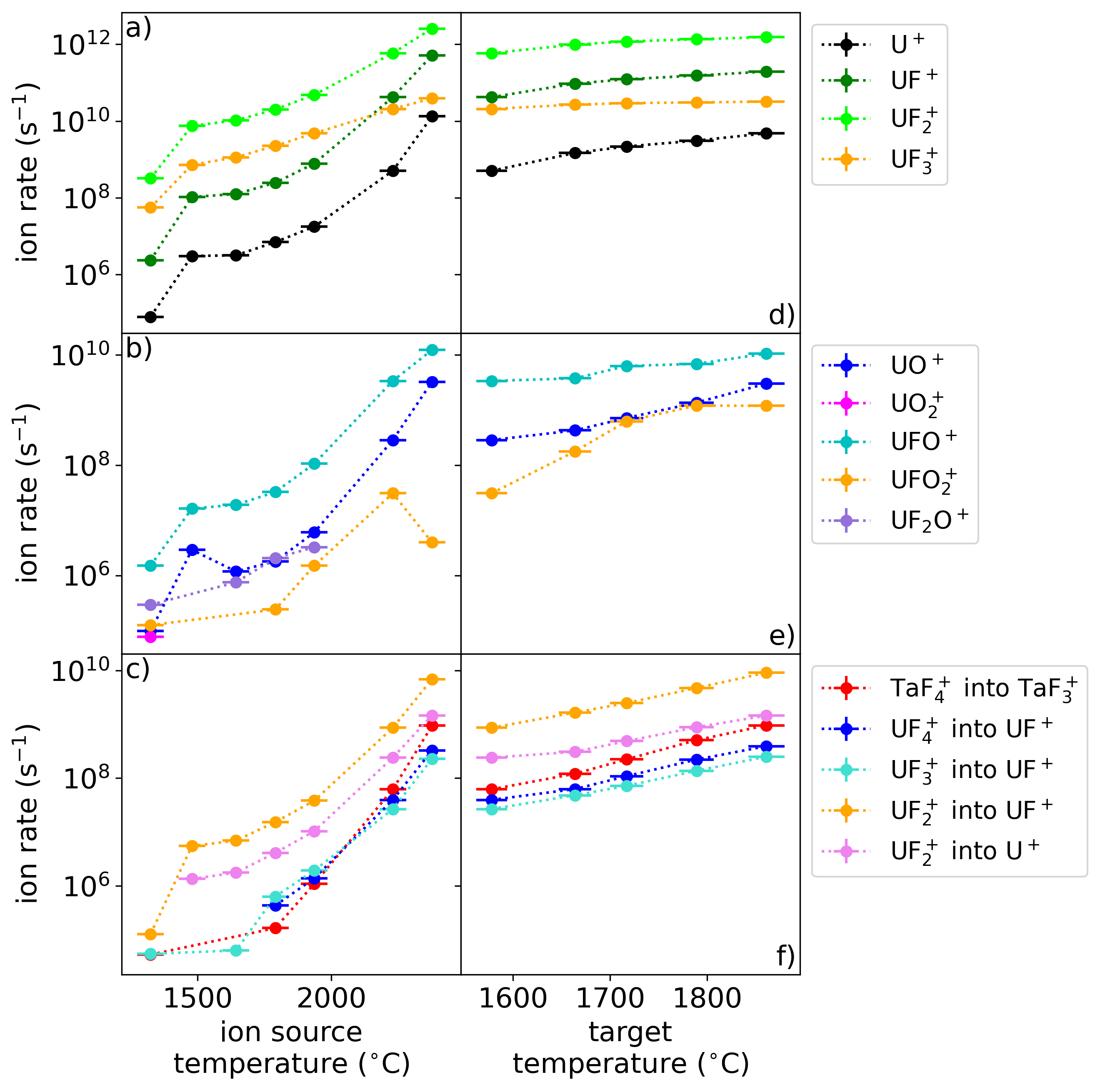}%
	\caption{Rates of ion beams from a tungsten surface ion source recorded on a Faraday Cup after mass separation from the ISOLDE separator magnets shown in logarithmic scale. a), b), c): as a function of ion-source temperature for a target temperature of 1578\,$^{\circ}$C. d) e) f): rates as a function of target temperature for an ion source temperature of 2230\,$^{\circ}$C. c) and f): rates of fragment ions measured on the apparent mass corresponding to the indicated dissociation of metastable molecular ions. \label{fig:linescan}}
\end{figure}

In mass spectrometry, the term ‘metastable’ is used to describe molecular ions possessing sufficient excess energy to fragment in the field-free region after leaving the ion source  \cite{Johnstone2012}. Upon fragmentation, the fragment ions retain a fraction of the kinetic energy of the extracted precursor ion. This causes fragment ions to pass through the mass separator magnetic field with an apparent mass $m^*$ corresponding to \cite{Johnstone2012}
\begin{equation}
	m^{*}=\frac{m_f^2}{m_p}
\end{equation}
where $m_f$ represents the mass of the fragment ion and $m_p$ represents the mass of the precursor metastable molecular ion. Fragment ions and their precursors were identified from the apparent mass and studied as a function of the target and surface ion source temperatures (Fig.~\ref{fig:linescan}). Increasing the ion source temperature significantly increased the fragment ion intensity, suggesting that at high temperatures, the molecules are more likely to have sufficient excess energy to reach the metastable states that fragment after extraction. Fragment molecules are indicated where observed in Fig.~\ref{fig:surface_vs_plasma}. 

\section{In-trap molecular formation}
\label{sec:in-trap}

\begin{figure}
	\includegraphics[width=0.5\textwidth]{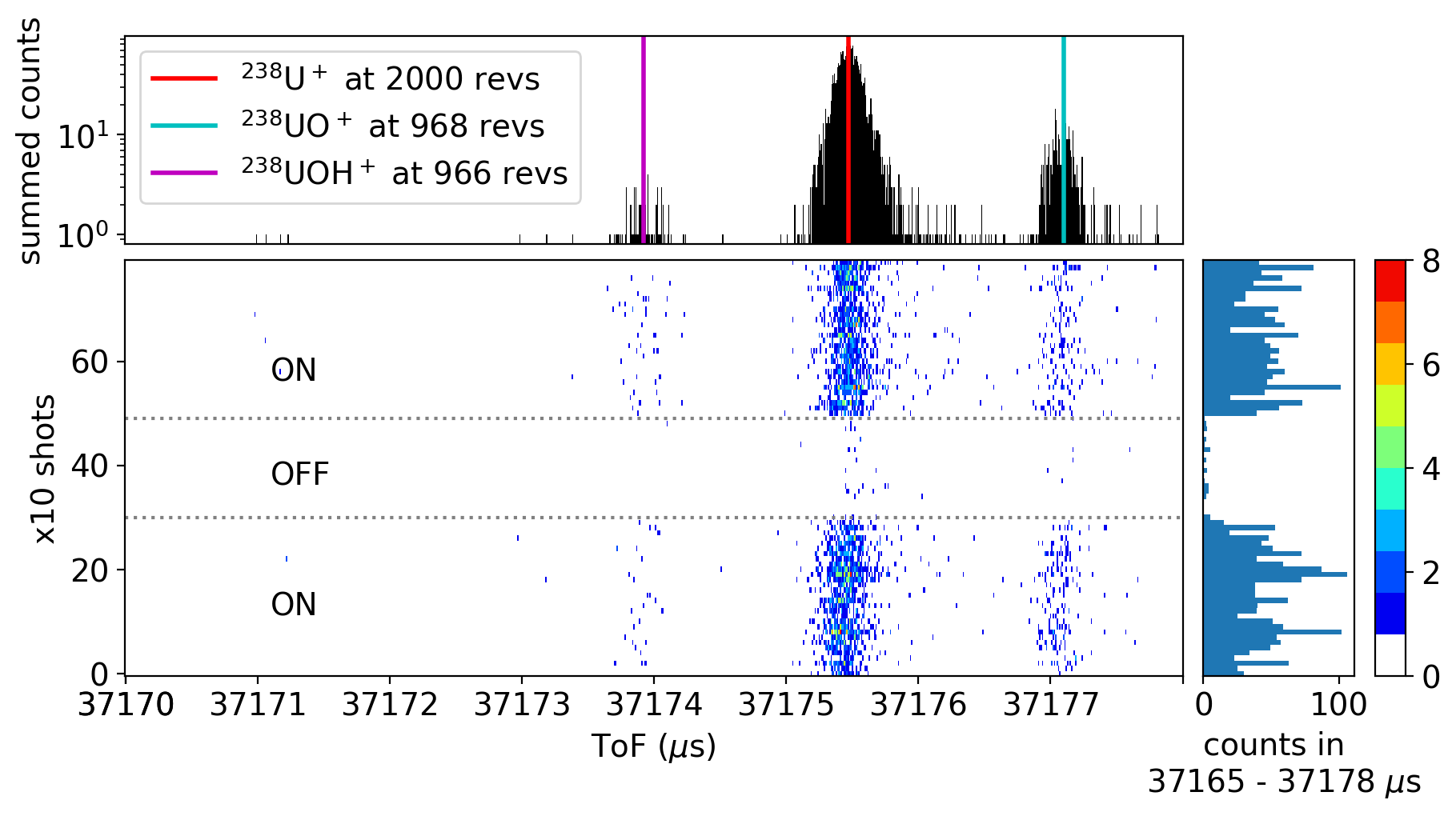}%
	\caption{ToF spectrum of A=238 mass-separated ion beams after cooling and bunching in the ISOLTRAP RFQ-cb, then trapping for 2000 revolutions in the MR-ToF MS as calculated for $^{238}$U. The status of the U resonant laser (on or off) is indicated. Vertical lines shown in the top panel indicate ToFs expected from offline calibrations. See text for further details. \label{fig:MRToFT154}}
\end{figure}

\begin{figure}
	\includegraphics[width=0.5\textwidth]{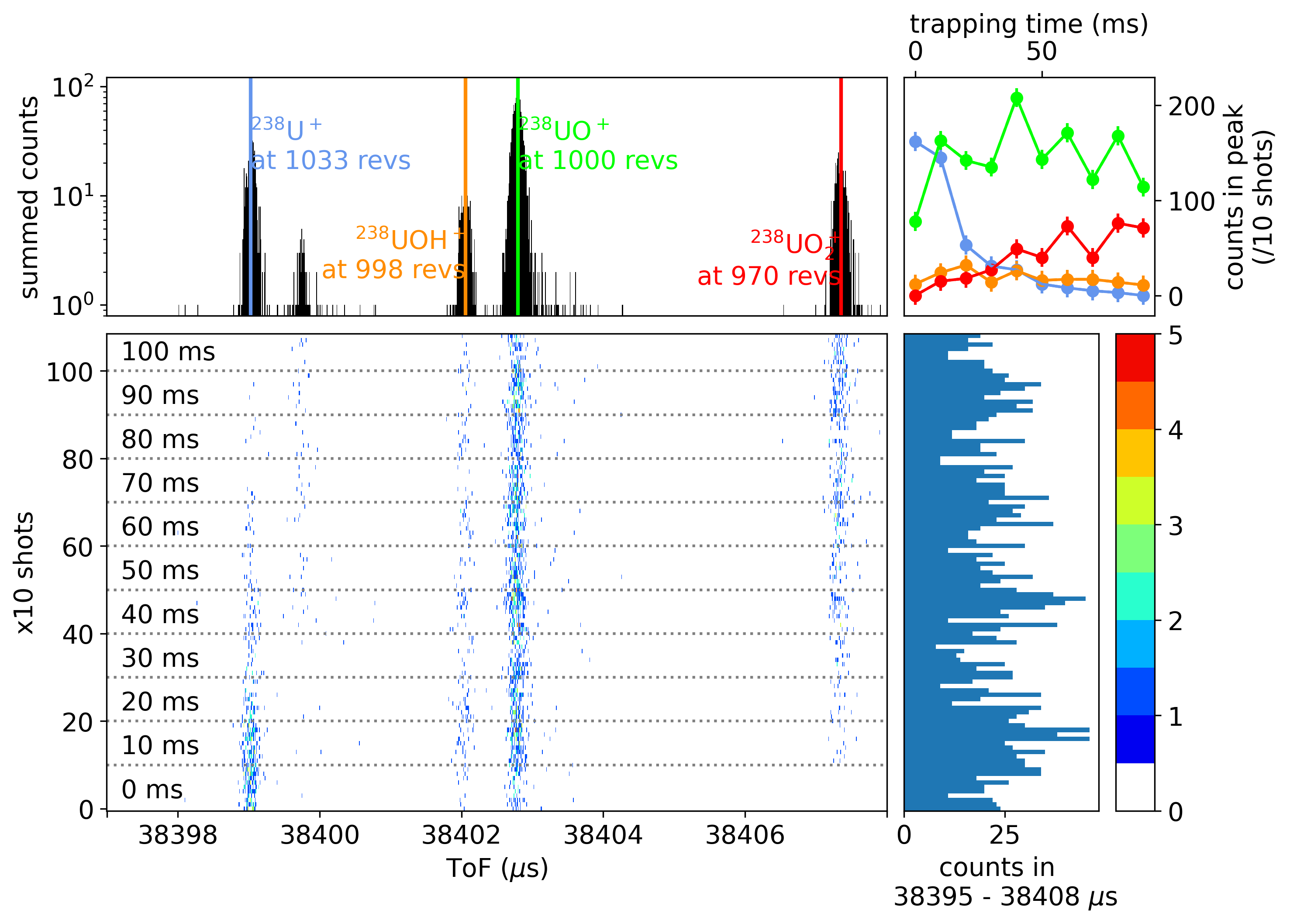}%
	\caption{ToF spectra for various storage times (indicated on the left) of A=238 mass-separated ion beams from the tungsten surface ion source with RILIS for U after cooling and bunching in the ISOLTRAP RFQ-cb, then trapping for 2000 revolutions in the MR-ToF MS as calculated for $^{238}$UO$^+$. Summed counts in each of the identified peaks are shown as a function of trapping time. See text for further details. \label{fig:TISDMOL_023}}
\end{figure}

\begin{figure}
	\includegraphics[width=0.45\textwidth]{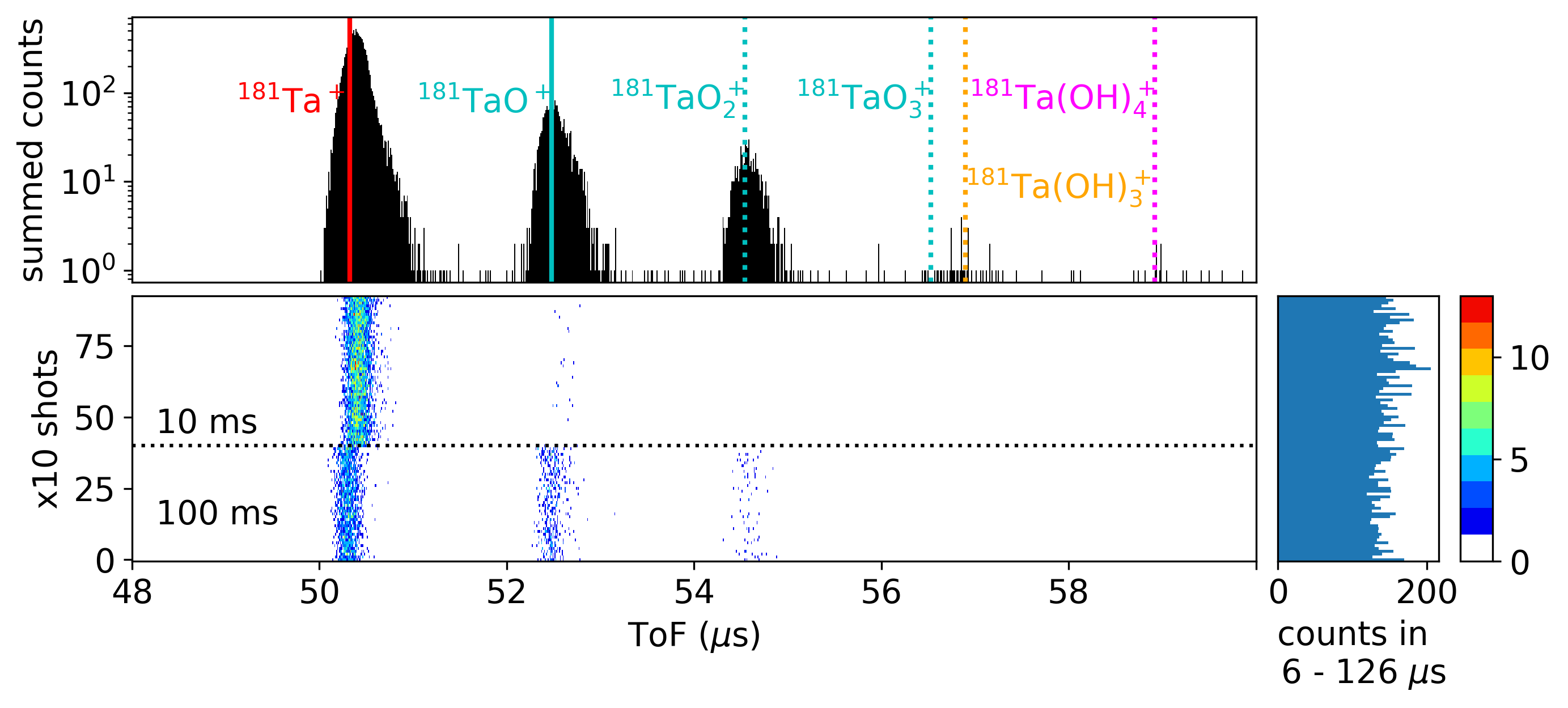}%
	\caption{ToF spectrum of A=181 mass-separated ion beams from the FEBIAD ion source after bunching in the ISOLTRAP RFQ-cb. Cooling time is indicated. Vertical lines in the top panel show ToFs expected from the offline calibration for single-pass operation of the MR-ToF MS. \label{fig:TISD119}}
\end{figure}

To study in-trap molecular formation, the ISOLTRAP RFQ-cb was employed with a buffer gas (here He at up to 10$^{-5}$\,mbar measured within 1\,m of the injection) to cool and bunch the ions. Mass-separated beams ionized using each of the studied ion sources were sent to the RFQ-cb for cooling and bunching. For studies of beam composition and in-trap molecular formation, a sample of the continuous ion beam was taken into the RFQ-cb. Ions were confined in the RFQ-cb for a trapping time during which interaction occurred between the ions, the buffer gas and residual gas contamination. The ion bunch was then ejected from the RFQ-cb and the arrival times of ions in each shot were measured with respect to the ejection time. Identification was performed with the MR-ToF MS using expected ToF values extracted from a calibration using $^{85,87}$Rb$^+$, $^{133}$Cs$^+$ from the ISOLTRAP offline ion source \cite{Wienholtz2013}, and online $^{238}$U$^+$ from ISOLDE. ToF spectra were accumulated over a number of shots as seen in Figures \ref{fig:MRToFT154}, \ref{fig:TISDMOL_023}, and \ref{fig:TISD119}.

\begin{figure}
	\includegraphics[width=0.5\textwidth]{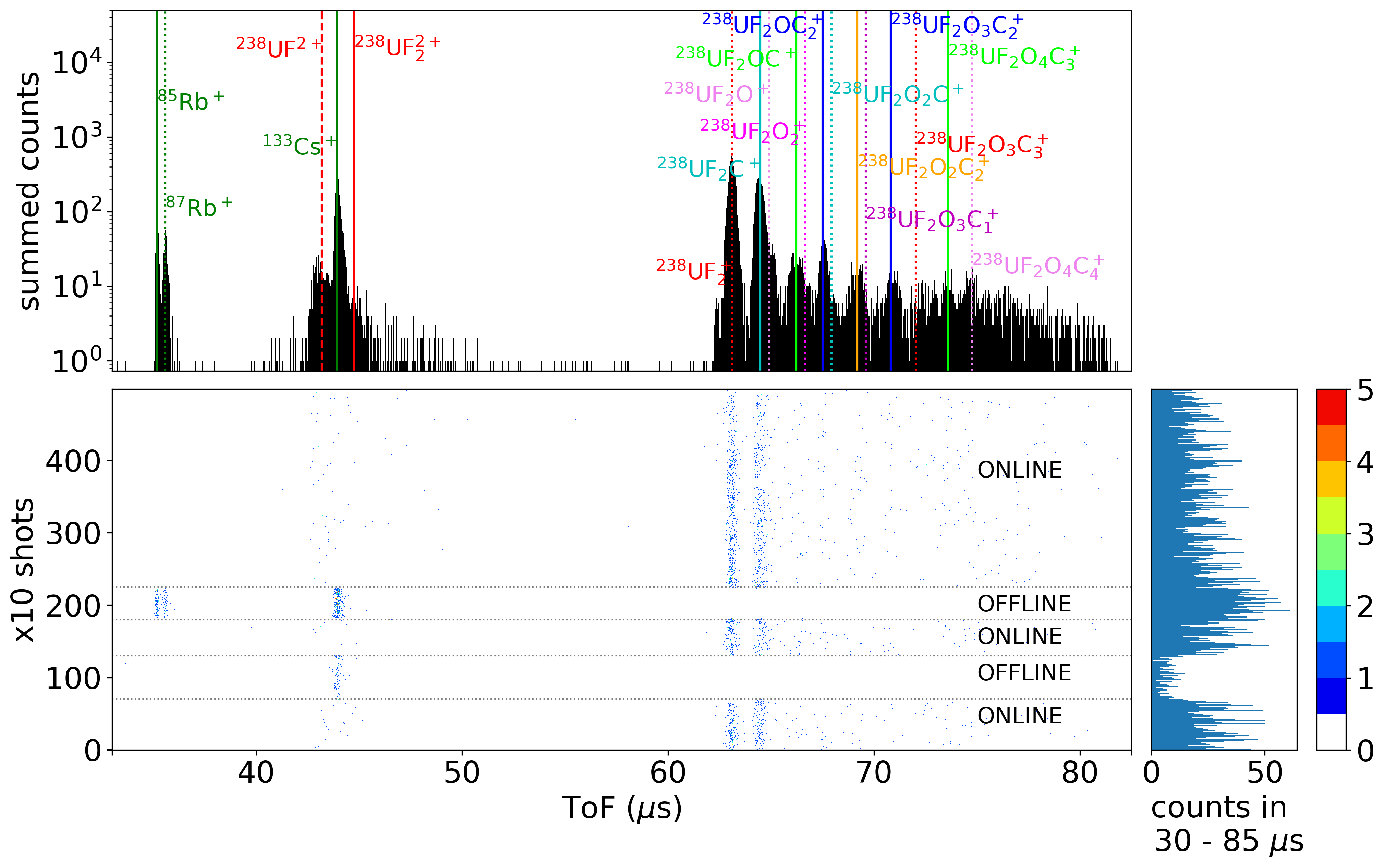}%
	\caption{ToF spectra of A=276 mass-separated ion beams from a W surface source with CF$_4$ injection in single-pass operation of the MR-ToF MS with trapping time 200 ms. Horizontal lines show switching between offline reference beam ($^{85,87}$Rb$^+$, $^{133}$Cs$^+$). Vertical lines show identified non-isobaric actinide molecules from molecular formation and charge exchange in the ISOLTRAP RFQ-cb. \label{fig:ToF-IDs}}
\end{figure}

The atomic uranium resonant laser ionization scheme in the ion source affected the count rates of uranium molecules (e.g. UO$^+$ and UOH$^+$ in Fig.~\ref{fig:MRToFT154}) observed from the RFQ-cb ion trap. Combined with the mass-separation step of the separator magnets, this indicates that the molecules are formed from ions in the mass-separated U$^+$ beam rather than in the target or ion source. For U and Ta, ratios depend on the trapping time as shown in Figures~\ref{fig:TISDMOL_023} and \ref{fig:TISD119}. In addition to atomic ions forming molecules, molecular ions mass-separated by ISOLDE (including UF$^+$, UF$_2^+$, TaF$^+$, TaF$_2^+$) reacted with the residual gas or buffer-gas contaminants to form molecules by pickup of C, O, H and OH (Figs.~\ref{fig:TISD119},\ref{fig:ToF-IDs}) and in some cases (UF$^+$, UF$_2^+$) were observed with higher charge states (UF$^{2+}$, UF$_2^{2+}$). To avoid detector saturation, attenuators were used to reduce the intensity of the ion beam injected into the RFQ-cb. This reduced absolute rates of in-trap formation and the formation efficiency relative to the ion beam intensity extracted from the ion source. Rates of molecular formation in the ion-trap represent ratios in the regime below space-charge limitations.

Notably, the in-trap UO$_x$ formation showed an identical response to the storage time in the RFQ-cb before and after the addition of a liquid nitrogen cold trap to the buffer gas line, indicating that reaction products enter the ion trap through diffusion into the vacuum chamber rather than buffer gas injection. 

\section{Conclusions}
\label{sec:conclusions}

For previously irradiated or oxidized targets, UO$^+$ and UO$_2^+$ sidebands are on the order of $10^5$\,ions\,s$^{-1}$ at target temperatures above 1600\,$^{\circ}$C. Without further addition of oxygen, the intensity of these contaminants will decrease over time, with the UO$_2^+$ sideband depleting first, followed by the UO$^+$ sideband. At nominal target temperatures (2000\,$^{\circ}$C) and above, UC$^+$, UC$_2^+$ can similarly reach rates of $10^4$ ions\,s$^{-1}$ or more. Fragments formed from the dissociation of metastable U and Ta fluorides arrive as additional non-isobaric contaminants in mass-separated beams. Fragment ions and their precursor ions can be identified using their apparent mass (Eq. 1) and anticipated for a given mass-to-charge ratio with the rates presented here. We present some representative rates of molecular ions extracted from different ion sources with oxidation, CF$_4$, and target temperatures noted in Figure~\ref{fig:linescan} and Table~\ref{tab:in-target_rates} as well as some representative rates for formation of molecular ions in the RFQ-cb with trapping times noted in Table~\ref{tab:in-trap_rates}. Rates from the ion source depend very strongly on target and ion source temperature and CF$_4$ injection rate. Rates from the RFQ-cb depend on trapping time. 
{\setstretch{1.0}
\begin{table} [h]
	\begin{tabular*}{0.475\textwidth}{@{\extracolsep{\fill}}l l l l}
		\hline
		Ion source & Target & Species & Rate \\ 
		 &  temperature & & (ions\,s$^{-1}$)\\
		 & ($^{\circ}$C )& &  \\
		\hline
		Re surface&1600 & $^{235}$U$^+$ & 2.5(1)\,E+3\\ 
		previously&1600 & $^{238}$U$^+$ & 8.9(6)\,E+5 \\ 
		irradiated&1600  & $^{235}$U$^{16}$O$^+$ & 4.0(6)\,E+3 \\
		&1600  & $^{238}$U$^{16}$O$^+$ & 1.4(2)\,E+6 \\
		&1600& $^{238}$U$^{18}$O$^+$ & 2.5(9)\,E+3 \\
		&1600  & $^{235}$U$^{16}$O$_2^+$ & 2.4(1)\,E+3 \\
		&1600 & $^{238}$U$^{16}$O$_2^+$ & 7.9(5)\,E+5 \\
		&1600 & $^{238}$U$^{16}$O$^{18}$O$^+$ & 3.5(5)\,E+3 \\
		\hline
		Re surface&2100 & $^{235}$U$^+$ & 6.3(1)\,E+5\\ 
		previously&2100 & $^{238}$U$^+$ & 5(1)\,E+7 \\ 
		irradiated&2100 & $^{235}$U$^{12}$C$^+$ & 7(1)\,E+3 \\
       &2100 & $^{235}$U$^{16}$O$^+$ & 1.24(6)\,E+4\\
       &2100 & $^{238}$U$^{16}$O$^+$ & 3.3(2)\,E+6 \\
		&2100 & $^{238}$U$^{18}$O$^+$ & 8(1)\,E+3 \\
		&2100 & $^{235}$U$^{12}$C$_2^+$ & 4.5(7)\,E+4  \\
		&2100 & $^{238}$U$^{12}$C$_2^+$ & 1.3(2)\,E+7 \\
		&2100 & $^{238}$U$^{12}$C$^{13}$C$_2^+$ & 3(2)\,E+5 \\
		&2100 & $^{238}$U$^{16}$O$_2^+$ & 7.3(3)\,E+4 \\
		&2100 & $^{238}$U$^{16}$O$^{18}$O$^+$ & 7.4(8)\,E+2 \\
		&2100 & $^{238}$U$^{12}$C$_3^+$ & 1.3(4)\,E+4 \\
		\hline
		FEBIAD&1818&$^{238}$U$^{16}$O$^{2+}$ & 2.8(5)\,E+7\\
		with&1818&$^{238}$U$^{16}$F$^{2+}$ & 6(1)\,E+8\\
		CF$_4$&1818&$^{238}$U$^+$ & 1.5(1)\,E+11\\
		&1818&$^{238}$U$^{16}$O$^+$ &3.2(2)\,E+11 \\
		&1818&$^{238}$U$^{19}$F$^+$ & 1.7(1)\,E+11\\
		&1818&$^{238}$U$^{19}$F$^{16}$O$^+$ & 1.3(1)\,E+10\\
		&1818&$^{238}$U$^{19}$F$_2^+$ & 1.4(1)\,E+11\\
		&1818&$^{238}$U$^{19}$F$^{16}$O$_2^+$ & 3.0(7)\,E+7\\
		&1818&$^{238}$U$^{19}$F$_2^{16}$O$^+$ & 3.5(5)\,E+9\\
		&1818&$^{238}$U$^{19}$F$_3^+$ & 6.8(5)\,E+11\\
		&1818&$^{238}$U$^{19}$F$_4^+$ & 4.9(4)\,E+10\\
		\hline
	\end{tabular*}
	\caption{Some observed rates of uranium ion beams from a rhenium surface source and from an electron impact ion source with injected CF$_4$, recorded on a Faraday cup after mass separation from the ISOLDE mass separator magnet.}
	\label{tab:in-target_rates}
\end{table}
}
{\setstretch{1.0}
\begin{table} [h]
	\begin{tabular*}{0.5\textwidth}{@{\extracolsep{\fill}}l l l l}
		\hline
		Trapping time & Initial ion & Species & Rate \\ 
		(ms) & beam & & (ions\,s$^{-1}$ of\\
		&  & &  injection time)\\
		\hline
		5&$^{234}$U$^+$ & $^{234}$U$^+$ & 8.0(8)\,E-2\\ 
		5&$^{234}$U$^+$ & $^{234}$UO$_2$H$^+$ & 9(1)\,E-2 \\ 
		5&$^{235}$U$^+$ & $^{235}$U$^+$ & 1.64(5)\\
		5&$^{235}$U$^+$ & $^{235}$UO$_2^+$ & 5(1)\,E-2\\
		5&$^{238}$U$^+$ & $^{238}$U$^+$ & 4.40(7)\,E+4\\ 
		5&$^{238}$U$^+$ & $^{238}$UO$^+$ & 5.6(2)\,E+2 \\ 
		5&$^{238}$U$^+$ & $^{238}$UOH$^+$ & 1.5(1)\,E+2\\
		
		5&$^{235}$UF$^+$ & $^{235}$UF$^+$ &  1.39(3)\,E+4\\
		5&$^{235}$UF$^+$ & $^{235}$UFO$^+$ &  3.6(3)\,E+2\\
		5&$^{235}$UF$^+$ & $^{238}$UFOH$^+$ &  7(1)\,E+1\\
		100&$^{235}$UF$^+$ & $^{235}$UFO$^+$ &  1.16(6)\,E+4\\
		100&$^{235}$UF$^+$ & $^{238}$UFOH$^+$ &  5.2(9)\,E+2\\
		100&$^{238}$UF$^+$ & $^{238}$UF$^+$ &  1.39(8)\,E+5\\		
		100&$^{238}$UF$_2^+$ & $^{238}$UF$_2^+$ &  1.69(3)\,E+4\\
		100&$^{238}$UF$_2^+$ & $^{238}$UF$_2$O$^+$ &  7(1)\,E+1\\
		100&$^{238}$UF$_2^+$ & $^{238}$UFO$_2$H$^+$ &  1.6(6)\,E+1\\
		200&$^{238}$UF$_2^+$ & $^{238}$UF$_2^+$ &  1.17(2)\,E+4\\
		200&$^{238}$UF$_2^+$ & $^{238}$UF$_2$O$^+$ &  3.4(3)\,E+2\\
		200&$^{238}$UF$_2^+$ & $^{238}$UFO$_2$H$^+$ &  9.2(1.4)\,E+1\\
		200&$^{238}$UF$_2^+$ & $^{238}$UF$^{2+}$ &  8.2(9)\,E+1\\
		200&$^{238}$UF$_2^+$ & $^{238}$UF$_2^{2+}$ &  2.9(5)\,E+1\\
		\hline
	\end{tabular*}
	\caption{Observed rates of ions formed inside the RFQ-cb from initial mass-separated ion beams extracted from a tungsten surface ion source. Counts per seconds of initial beam sampling time were recorded on the ISOLTRAP MagneToF detector after mass separation in the MR-ToF MS. See text for more details.}
	\label{tab:in-trap_rates}
\end{table}
}

Production of fluoride molecular ions from the ion source is achieved by adding CF$_4$. Since many actinide fluorides are stable at temperatures above 1000\,$^{\circ}$C, in-source formation is a promising approach that can use parameters including target and ion source temperatures and fluorine partial pressure to control formation rates. To form molecules that may not be stable at high temperatures, molecular formation in the RFQ ion trap is presented as a possible approach. Formation of oxides, carbides, and hydroxides from the mass-separated atomic and molecular ion beams occurs in the ISOLTRAP RFQ-cb in the presence of residual gases. Trapping time is shown to be a parameter influencing the formation of molecules in the ion trap. The molecular formation reported here in the ISOLTRAP RFQ-cb may have implications for other RFQ-cb ion traps used in beam preparation (e.g. the ISOLDE cooler ISCOOL~\cite{Catherall2017}) which require further investigations.

These studies combined characterize the composition of beams heavier than the target material and provide information on the process of creating molecular actinide beams in targets and in ion traps. 

\section{Acknowledgements}
The authors gratefully acknowledge support from the ISOLDE operations team, the ISOLDE targets and ion sources team, and Simone Gilardoni. This project has received funding from the European Union's Horizon 2020 Research and Innovation Program (grant No. 861198 project ‘LISA’ MSC ITN). The authors acknowledge support from the German Federal Ministry of Education and Research (BMBF) for ISOLTRAP (grant No. 05P18HGCIA and 05P21HGCI1). L.N. acknowledges support from the Wolfgang Gentner Program (grant No. 13E18CHA).



\bibliographystyle{elsarticle-num} 
\bibliography{bib.bib}





\end{document}